\documentclass[9pt,conference]{IEEEtran}

\usepackage{waspaa25} 

\usepackage{bm} 
\usepackage{multirow} 

\title{Modeling Multi-Level Hearing Loss for Speech Intelligibility Prediction}

\name{XIAJIE ZHOU\thanks{This work was supported by JSPS KAKENHI Grant Numbers (20KK0233, 21H03463, 25H01139 and 25K21245).},
      CANDY OLIVIA MAWALIM,
      MASASHI UNOKI}
\address{Japan Advanced Institute of Science and Technology, Nomi, Japan
}

\begin{document}

\maketitle

\begin{abstract}
The diverse perceptual consequences of hearing loss severely impede speech communication, but standard clinical audiometry, which is focused on threshold-based frequency sensitivity, does not adequately capture deficits in frequency and temporal resolution. To address this limitation, we propose a speech intelligibility prediction method that explicitly simulates auditory degradations according to hearing loss severity by broadening cochlear filters and applying low-pass modulation filtering to temporal envelopes. Speech signals are subsequently analyzed using the spectro-temporal modulation (STM) representations, which reflect how auditory resolution loss alters the underlying modulation structure. In addition, normalized cross-correlation (NCC) matrices quantify the similarity between the STM representations of clean speech and speech in noise. These auditory-informed features are utilized to train a Vision Transformer–based regression model that integrates the STM maps and NCC embeddings to estimate speech intelligibility scores. Evaluations on the Clarity Prediction Challenge corpus show that the proposed method outperforms the Hearing-Aid Speech Perception Index v2 (HASPI v2) in both mild and moderate-to-severe hearing loss groups, with a relative root mean squared error reduction of 16.5\% for the mild group and a 6.1\% reduction for the moderate-to-severe group. These results highlight the importance of explicitly modeling listener-specific frequency and temporal resolution degradations to improve speech intelligibility prediction and provide interpretability in auditory distortions.
\end{abstract}

\section{Introduction}
Hearing loss is one of the most prevalent chronic health conditions globally, with significant impacts on communication, education, and quality of life \cite{patrick2022hearing}. Recent epidemiological analyses have highlighted a growing burden of hearing impairment in all age groups and regions, with hearing loss projected to affect more than 630 million individuals by 2030 and nearly 900 million by 2050 if no effective intervention is implemented \cite{li2022prevalence, wilson2017global}. Hearing loss is not a uniform condition: it varies widely in severity, ranging from mild to profound. These deficits result in reduced frequency selectivity and impaired temporal resolution, both of which degrade the listener’s ability to comprehend speech.

Speech intelligibility—the degree to which a listener correctly recognizes speech—directly reflects the listener’s ability to comprehend speech, especially when hearing is degraded. Speech intelligibility is thus a key outcome for evaluating real-world listening performance in hearing aid usage \cite{miles2022measuring}. While audiograms are widely used to measure hearing thresholds at different frequencies, they often fall short in explaining why some people understand speech better than others \cite{ching1998speech, smoorenburg1992speech}. Even listeners with similar audiograms can perform quite differently, especially in noisy situations \cite{pienkowski2017etiology}. This suggests that evaluating hearing loss should go beyond thresholds and also consider how well the ear processes sounds, particularly its ability to distinguish frequencies and follow rapid changes over time.


Although audiometric thresholds provide a basic estimate of hearing sensitivity, they fail to capture the auditory abilities essential for speech perception. One such ability is frequency resolution—the capacity to distinguish adjacent spectral components of sound—which is especially important in noisy environments. Impaired spectral resolution, which is commonly associated with sensorineural hearing loss, has been shown to significantly degrade intelligibility even when audibility is restored through amplification \cite{jin2010interrupted}. Simulation studies have further demonstrated that broader auditory filters based on individual audiograms lead to substantial reductions in speech understanding in complex acoustic scenes \cite{porhun2021method}. These findings indicate that speech intelligibility prediction must consider hearing thresholds and the extent of auditory filter degradation.

In addition to frequency resolution, temporal resolution—the ability to track rapid amplitude fluctuations—is essential for speech intelligibility \cite{haggard1985temporal}. These modulations convey critical cues related to voicing, syllabic timing, and phoneme boundaries. Temporal resolution is often measured using the temporal modulation transfer function (TMTF), which reflects a listener’s sensitivity to amplitude modulation at various rates. Research indicates that individuals with sensorineural hearing loss commonly show reduced modulation sensitivity and lower cutoff frequencies, especially at higher modulation rates \cite{moore1992temporal, bacon1992modulation}. These deficits vary considerably across listeners. Some maintain relatively intact temporal processing, while others—particularly those with more severe or sloping high-frequency losses—exhibit pronounced reductions in TMTF sensitivity \cite{desloge2011temporal}. This variability underscores the importance of accounting for individual differences in temporal resolution when predicting speech intelligibility.

\begin{figure*}[t]
\centering
\includegraphics[width=0.786\textwidth]{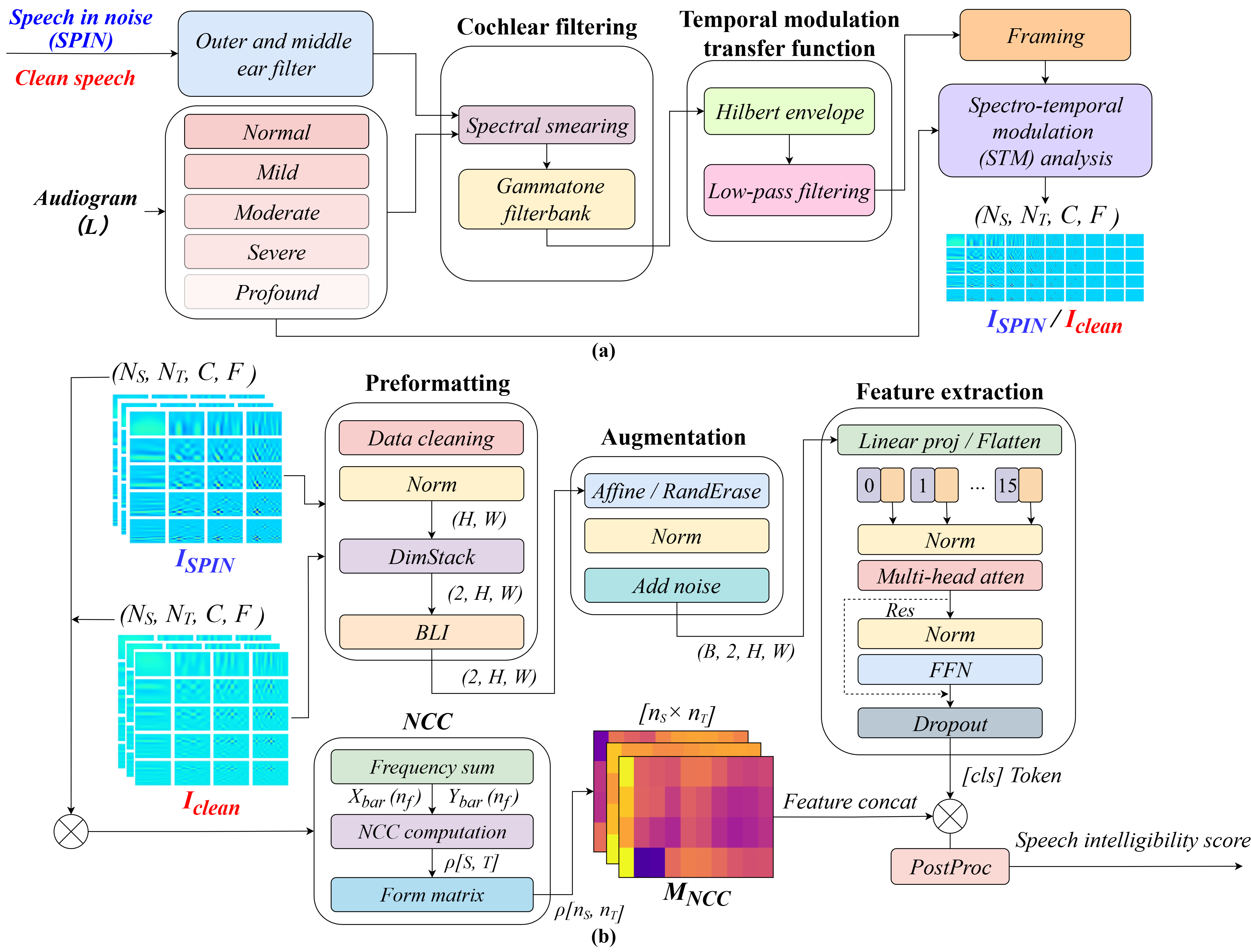}
\vspace{-5pt}
\caption{Overview of proposed method. \textbf{(a)}  STM representations are extracted from both clean speech and speech in noise (SPIN) after peripheral processing including inverse equal-loudness contour (ELC) correction, producing 4D tensors $\mathbf{I}_{\mathrm{clean}}$ and $\mathbf{I}_{\mathrm{SPIN}}$ with dimensions $(N_S, N_T, C, F)$, where $N_S$ and $N_T$ are the number of spectral and temporal modulation channels, $C$ is the number of cochlear frequency bands, and $F$ is the number of time frames. \textbf{(b)} The two STM tensors are preprocessed into stacked image inputs via normalization (Norm), bilinear resizing (BLI), and dimensional stacking (DimStack). Data augmentation (e.g., affine transform, random erasing, and noise injection) is optionally applied. In parallel, normalized cross-correlation (NCC) is computed between clean speech and SPIN STM representations to generate a modulation similarity matrix $\mathbf{M}_{\mathrm{NCC}} \in \mathbb{R}^{N_S \times N_T}$. Finally, a Vision Transformer (ViT) backbone based on the ViT-Base architecture receives the STM representations and the flattened NCC matrix. The transformer’s classification token ($\mathrm{[CLS]}\ \mathrm{Token}$) is concatenated with the NCC embedding and passed through a regression head to predict the final speech intelligibility score.}
\label{fig:method_vitncc}
\end{figure*}

Since spectral and temporal resolution are degraded under hearing loss, the ability to extract dynamic acoustic cues is compromised. However, these two domains are not processed in isolation. Natural speech exhibits spectro-temporal modulations—joint fluctuations in frequency and time—that are believed to be encoded by the auditory cortex in the auditory system \cite{chi1999spectro}. We adopt a spectro-temporal modulation (STM) analysis based on two-dimensional Gabor filters to capture this integrated structure \cite{edraki2020speech}. This biologically inspired framework mirrors cortical tuning patterns and provides a unified representation of modulation content, allowing us to assess how hearing loss alters the encoding of speech-relevant modulations.

This study aims to predict speech intelligibility under hearing loss by explicitly accounting for listener-specific degradations in frequency and temporal resolution. (1) We simulate the effects of hearing loss using auditory-inspired transformations that reflect varying degrees and types of spectral and temporal deficits. (2) We characterize speech signals through STM analysis, which captures how these degradations distort the modulation patterns critical for speech intelligibility. Unlike prior methods relying solely on audiometric thresholds or broadband metrics, our approach generates a time–frequency–modulation representation tailored to the listener's perceptual limitations. These STM representations are then utilized to train a regression model for prediction.

\section{Related work}
The Hearing-Aid Speech Perception Index v2 (HASPI v2) is adopted as the baseline intrusive model in the Clarity Prediction Challenge \cite{kates2021hearing, kates2022overview, barker2024clarity}. It simulates impaired auditory processing using the Gammatone filterbank, dynamic range compression, and envelope extraction. Hearing loss effects are incorporated via audiogram-based attenuation and reduced frequency selectivity, with speech intelligibility estimated by correlating clean speech and speech in noise (SPIN) envelopes across time–frequency bands.

To account for binaural listening, HASPI computes left and right ear scores and selects the higher one (\textit{HASPI v2}). A logistic function is applied to map HASPI scores to the final prediction:
\begin{equation}
\hat{y} = \frac{1}{1 + e^{-a(\text{HASPI} - b)}},
\end{equation}
where $a$ and $b$ are optimized scaling parameters, and $\hat{y} \in [0, 1]$ denotes the predicted score.

\section{Proposed method}
\begin{table*}[t]
\centering
\caption{Frequency resolution simulation settings for each hearing loss level, based on Nejime et al. (1997) and the implemented auditory filter model.}
\label{tab:freqsettings}
\begin{tabular}{lcccc}
\toprule
\textbf{Hearing Loss Level} & \textbf{Filter Broadening (ERB)} & \textbf{Recruitment} & \textbf{Simulation Description} & \textbf{No. of Channels} \\
\midrule
Normal Hearing & 1.0× ERB & None & No distortion in filter shape or gain & 36 \\
Mild Loss      & 1.5× ERB & Mild expansion & Slight filter broadening and mild compression & 36 \\
Moderate Loss  & 2.0× ERB & Moderate expansion & Moderately impaired frequency selectivity & 28 \\
Severe Loss    & 3.0× ERB & Strong expansion & Strong filter smearing with loudness recruitment & 19 \\
\bottomrule
\end{tabular}
\end{table*}

\begin{table}[t]
\centering
\caption{Time constant $\tau$ and cutoff frequency $f_{3\mathrm{dB}}$ for each hearing condition.}
\label{tab:tmftparams}
\begin{tabular}{lcc}
\toprule
\textbf{Hearing Loss Level} & $\tau$ (ms) & $f_{3\mathrm{dB}}$ (Hz) \\
\midrule
Normal Hearing & 2.2 & 72.3 \\
Mild Loss      & 3.4 & 46.8 \\
Moderate Loss  & 5.0 & 31.8 \\
Severe Loss    & 9.4 & 16.9 \\
\bottomrule
\end{tabular}
\end{table}

A comprehensive overview of proposed method is shown in Fig. \ref{fig:method_vitncc}.
\subsection{Frequency Resolution Modeling in Auditory Periphery}
We utilize the MSBG hearing loss model based on Nejime's research, which plays a crucial role in adjusting audio signals to reflect the anticipated hearing loss effects \cite{nejime1997simulation, moore1999loudness}. The model relies on audiograms that represent the hearing thresholds of the listener to simulate individual hearing loss effects.

\textbf{Outer and Middle Ear Processing}: The initial stage of the model involves simulating the effects of the outer and middle ear. The model applies filters that replicate the acoustic transfer from the free field to the eardrum and through the middle ear. These filters are selected based on the physical characteristics of the auditory system and are independent of the audiogram. The signal first undergoes filtering to simulate the outer and middle ear's response, ensuring the signal is correctly attenuated or amplified before reaching the cochlea.

\textbf{Cochlear Filtering and Hearing Loss Simulation}: The cochlear stage of the model simulates three key effects of sensorineural hearing loss \cite{moore1999loudness}: (1) elevated hearing thresholds, (2) broadened auditory filters indicating reduced frequency selectivity, and (3) abnormal loudness growth (recruitment). After outer and middle ear filtering, the signal is passed through the Gammatone filterbank that mimics cochlear frequency decomposition. The filter bandwidths are adaptively broadened depending on hearing loss severity (i.e., equivalent rectangular bandwidth (ERB) \cite{moore1997model}), as summarized in Table~\ref{tab:freqsettings}, to reflect impaired frequency resolution. For each channel, loudness recruitment is modeled by an expansive nonlinearity applied to the envelope, simulating the reduced dynamic range in impaired listeners. Together, these adjustments produce frequency-specific signal degradations that represent different degrees of hearing impairment and are critical for modeling their effects on speech intelligibility.

\subsection{Temporal Resolution Modeling Based on Hearing Loss}
The choice of time constants $\tau$ was guided by empirical measurements of TMTF in both normal-hearing and hearing-impaired listeners~\cite{bacon1992modulation, desloge2011temporal}. These prior studies showed that TMTF—describing modulation sensitivity across modulation frequencies—can be well approximated by a first-order low-pass filter, capturing the decline in temporal resolution with increasing hearing loss.

We implemented this using a filter with an impulse response:
\begin{equation}
h(t) = \frac{1}{\tau} e^{-t/\tau} u(t),
\end{equation}
and magnitude frequency response:
\begin{equation}
|H(f)| = \frac{1}{\sqrt{1 + (2\pi f \tau)^2}}.
\end{equation}

The cutoff frequency at which $|H(f)|$ drops to $1/\sqrt{2}$ is given by
\begin{equation}
f_{\mathrm{cutoff}} = \frac{1000}{2\pi \tau},
\end{equation}
where $h(t)$ is the impulse response, $\tau$ is the time constant (ms), $u(t)$ is the unit step function, $H(f)$ is the magnitude response, $f$ is the modulation frequency (Hz), and $f_{\mathrm{cutoff}}$ is the $3$~$\mathrm{dB}$ cutoff frequency indicating the listener's upper limit for tracking envelope modulations.

Above this frequency, sensitivity declines sharply, impairing the perception of voicing, rhythm, and phoneme boundaries. As speech intelligibility relies heavily on modulation cues below $50$~$\mathrm{Hz}$, accurate modeling of this cutoff is essential.

We adopted the $\tau$ values reported in~\cite{desloge2010speech, desloge2011temporal, bacon1992modulation}, where first-order filters were fitted to TMTF data across different hearing loss levels. These values, which are summarized in Table~\ref{tab:tmftparams}, provide a physiologically motivated mapping from hearing condition to the temporal filtering parameters used in our simulation.

\subsection{Spectro-Temporal Modulation Analysis}
This study employs the STM analysis based on two-dimensional Gabor filters to extract stable and perceptually relevant modulation features from the speech envelope \cite{edraki2020speech,chi2005multiresolution,schadler2015separable}. The STM representations have been shown to effectively characterize critical modulation patterns for human speech understanding \cite{ponsot2021mechanisms}. STM analysis captures the joint spectro-temporal structure of speech, and prior studies have indicated that energy concentrated in specific modulation ranges—such as spectral modulation between 0.25--1 cycles/octave and temporal modulation between 2--16 Hz—is particularly predictive of speech intelligibility \cite{bernstein2013spectrotemporal}. Therefore, STM representations are both physiologically plausible and informative for prediction.

To simulate the auditory system’s selectivity to various modulation rates, the basic form of a Gabor filter is expressed as
\begin{equation}
g(t) = w(t)\cos(\omega t + \phi),
\end{equation}
where $w(t)$ is a windowed envelope, $\omega$ is the center modulation frequency, and $\phi$ is the phase. A Hann window is used as $w(t)$ to control filter locality. After construction, each Gabor filter is normalized so that its maximum frequency-domain magnitude equals $1.0$, ensuring consistency in energy representation across modulation channels \cite{schadler2015separable}.

Given that frequency and temporal resolution vary with hearing loss, we generate modulation center frequencies dynamically rather than using a fixed grid. Specifically, the center modulation frequency $\omega_i$ is computed recursively as
\begin{equation}
\omega_{i} = \frac{\omega_{\mathrm{max}}}{r^i}\quad \text{with} \quad \omega_{\mathrm{min}} = \frac{\pi\nu}{\mathrm{size}_{\mathrm{max}}},
\end{equation}
where $\omega_{\mathrm{max}}$ and $\omega_{\mathrm{min}}$ denote the upper and lower bounds of the modulation frequency range, respectively. The parameter $r$ determines the logarithmic spacing between adjacent filters, $\nu$ controls the number of half-wavelengths within the envelope (thus affecting filter bandwidth), and $\mathrm{size}_{\mathrm{max}}$ is the maximum extent of the envelope along the corresponding axis—i.e., the number of frequency channels for spectral filtering or the number of frames for temporal filtering.

This formulation ensures that the filters remain localized within the input envelope dimensions, thereby avoiding oversized or misaligned filters that would otherwise span outside the valid region.

Each signal is transformed into a four-dimensional STM tensor:
\begin{equation}
\mathrm{STM}[S, T, c, f] \in \mathbb{R}^{N_S \times N_T \times C \times F},
\end{equation}
where $N_S$ and $N_T$ denote the numbers of spectral and temporal modulation channels, $C$ is the number of auditory frequency bands, and $F$ is the number of time frames. This representation encodes modulation-specific energy distributions over time and frequency.

\subsection{Speech Intelligibility Prediction Model}
To estimate sentence-level speech intelligibility, we utilize the STM representations and the corresponding normalized cross-correlation (NCC) matrix as model inputs \cite{edraki2020speech, edraki2021spectro}. The NCC matrix $\mathbf{M}_{\mathrm{NCC}} \in \mathbb{R}^{N_S \times N_T}$ captures the modulation similarity between clean speech and SPIN across spectro-temporal subbands. For each modulation channel $(S, T)$, we first compute frequency-aggregated trajectories $X_{\mathrm{bar}}[n; S,T]$ and $Y_{\mathrm{bar}}[n; S,T]$ by summing along the frequency axis. The similarity is then measured using the Pearson correlation:
\begin{equation}
    \rho[S, T] = \frac{\langle X_{\mathrm{bar}} - \mu_X, \; Y_{\mathrm{bar}} - \mu_Y \rangle}{\|X_{\mathrm{bar}} - \mu_X\| \cdot \|Y_{\mathrm{bar}} - \mu_Y\|},
\end{equation}
where $\mu_X$ and $\mu_Y$ are the means of the respective trajectories. This yields a compact similarity map reflecting each modulation band's envelope degradation.

The STM representations (with clean and noisy speech as two channels) and the flattened NCC matrix are jointly used for predicting speech intelligibility. We adopt the Vision Transformer (ViT) architecture, specifically, the ViT-Base architecture \cite{dosovitskiy2020image}, which consists of 12 Transformer blocks, 12 attention heads, and a 768-dimensional embedding space\footnote{\url{https://github.com/google-research/vision_transformer}}. A trainable classification token ([CLS] token) is prepended to the sequence of STM patch embeddings to aggregate global visual information. In parallel, the NCC similarity matrix is projected into a low-dimensional embedding and concatenated with the [CLS] token before being passed to the regression head. This design follows the standard ViT paradigm, where the [CLS] token serves as a global representation and is further enhanced by incorporating modulation-based cross-signal similarity \cite{dosovitskiy2020image, liu2025metacls}. The final regression head outputs normalized scores in the range of $[0, 1]$.


\section{Experiment}
\subsection{Dataset}
We evaluated our proposed speech intelligibility prediction framework using the Clarity Prediction Challenge corpus\footnote{\url{https://claritychallenge.org/docs/cpc2/cpc2_data}}, which provides paired speech recordings and speech intelligibility scores collected from hearing-impaired listeners using simulated hearing aids. Each utterance consists of clean speech, SPIN, and the listener’s audiogram \cite{barker2024clarity}. 

Although the original dataset consists of unified training and testing partitions, we further stratified the data according to the degree of hearing loss. Specifically, using the left-ear pure-tone average, we categorized each listener into either the \textbf{Mild} group or the \textbf{Moderate-to-severe} group. This enables targeted analysis of how speech intelligibility prediction varies with hearing loss severity.

Table~\ref{tab:dataset_split} summarizes the number of utterances used in our training and evaluation. Note that seven utterances were excluded from the dataset due to SPIN signals lacking valid information. These signals exhibited severe degradation, resulting in unintelligible or blank outputs that prevented effective processing.

\begin{table}[t]
\centering
\caption{Utterance distribution by hearing loss severity, grouped by left-ear pure-tone average}
\label{tab:dataset_split}
\begin{tabular}{lcc}
\toprule
\textbf{Hearing Loss Group} & \multicolumn{2}{c}{\textbf{No. of Utterances}} \\
\cmidrule(lr){2-3}
                             & \textbf{Train} & \textbf{Test} \\
\midrule
Mild                        & 627            & 67            \\
Moderate-to-severe          & 11,609         & 830           \\
\midrule
\textbf{Total (used subset)} & \textbf{12,236} & \textbf{897}  \\
\bottomrule
\end{tabular}
\end{table}

\subsection{Experimental Settings}
We trained the proposed speech intelligibility model separately on two listener groups: \textbf{Mild} and \textbf{Moderate-to-severe}. Although the model architecture remained identical across both conditions, the training strategies differed slightly.

We employed a fixed training configuration for the \textbf{Mild} group, corresponding to a smaller dataset. The model was trained for 60 epochs with a batch size of 16, using STM representations of size $224 \times 224$ with two channels (clean speech and SPIN). The corresponding NCC matrix was flattened and projected into a 64-dimensional feature vector. Optimization was performed using AdamW with a learning rate of $1 \times 10^{-5}$ and a weight decay of $1 \times 10^{-4}$.

For the \textbf{Moderate-to-severe} group, we performed automatic hyperparameter tuning using Optuna \cite{akiba2019optuna}, exploring variations in learning rate, weight decay, dropout rate, and drop-path rate. The best model for each group was selected based on the lowest validation root-mean-square error (RMSE). 

As evaluation metrics, we used Pearson correlation coefficient ($\rho$) to assess linear correlation and RMSE to quantify differences.

\section{Results and Discussion}
We evaluated the model performance on the Clarity Prediction Challenge corpus, which was stratified into subgroups based on the degree of hearing loss. As both HASPI v2 and the proposed method are intrusive intelligibility metrics, they rely on access to the clean speech and SPIN at the hearing-aid output. The HASPI v2 used in this study corresponds to the better ear, which selects the higher score from the left and right ear simulations. In contrast, our proposed method utilizes only the left-ear signals.

\begin{figure}[t]
  \centering
  \centerline{\includegraphics[width=\columnwidth]{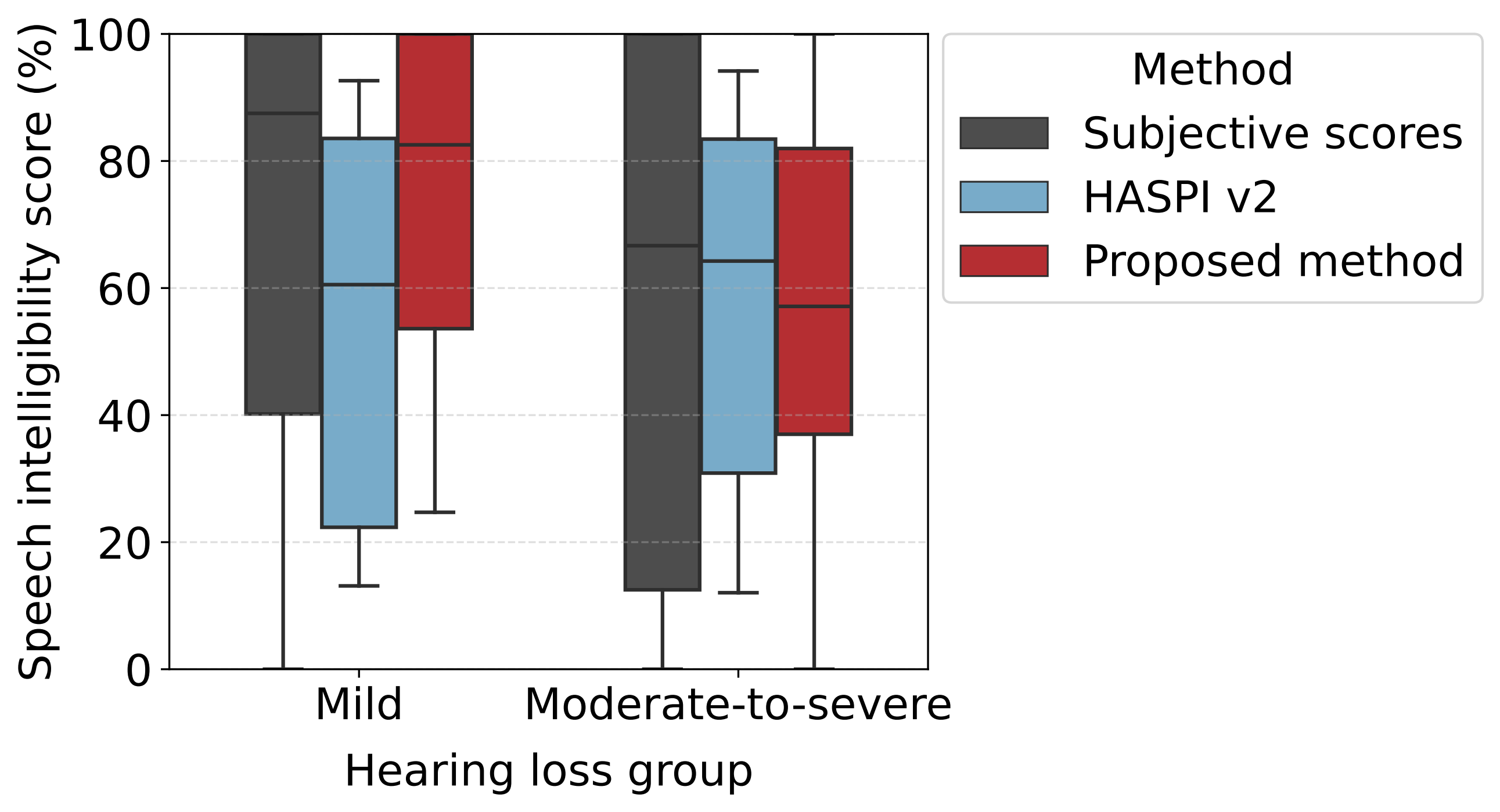}}\
  \vspace{-15pt}
  \caption{Distribution of predicted scores (HASPI v2 and proposed method) and subjective scores across \textbf{Mild} and \textbf{Moderate-to-severe} hearing loss groups. The proposed method shows improved alignment with subjective scores, particularly for mild loss listeners.}
  \label{fig:intelligibility_boxplot}
\end{figure}

Although both HASPI v2 and the proposed method output predicted scores in the normalized range of $[0,1]$, all evaluations were performed with respect to the subjective speech intelligibility scores, which range from $[0,100]$. Therefore, all predictions were linearly scaled to match the target range before computing evaluation metrics. Figure~\ref{fig:intelligibility_boxplot} shows the distribution of predicted scores across hearing loss groups after rescaling, highlighting the closer alignment between the proposed method and subjective ratings.

Table~\ref{tab:results_all} shows that the proposed method outperforms HASPI v2 both RMSE and $\rho$. Specifically, for the \textbf{Mild} group, the RMSE is reduced from 29.85 to 24.91, corresponding to an approximate \SI{16.5}{\percent} relative improvement in prediction accuracy. The Pearson correlation increases from 0.70 to 0.77, indicating more substantial alignment with subjective speech intelligibility scores. 

For the \textbf{Moderate-to-severe} group, the proposed method also shows improved $\rho$ (0.75 vs. 0.69), with slightly lower RMSE (27.04 vs. 28.79), suggesting robust performance even under more challenging hearing conditions. These results suggest that incorporating modulation-informed representations and correlation-based similarity enables better modeling for listeners with hearing loss, even when relying solely on monaural input.

\begin{table}[t]
\centering
\caption{Performance comparison between HASPI v2 and the proposed method across hearing loss groups.}
\label{tab:results_all}
\begin{tabular}{llcc}
\toprule
\textbf{Group} & \textbf{Method} & \textbf{RMSE} $\downarrow$ & \boldmath$\rho$ $\uparrow$ \\
\midrule
\multirow{2}{*}{Mild} 
  & HASPI v2~\cite{kates2021hearing} & 29.85 $\pm$ 1.49 & 0.70 \\
  & Proposed method               & \textbf{24.91 $\pm$ 1.25} & \textbf{0.77} \\
\midrule
\multirow{2}{*}{Moderate-to-severe} 
  & HASPI v2~\cite{kates2021hearing} & 28.79 $\pm$ 1.44 & 0.69 \\
  & Proposed method               & \textbf{27.04 $\pm$ 1.35} & \textbf{0.75} \\
\bottomrule
\end{tabular}
\end{table}

\section{Conclusion}
In this paper, we proposed a speech intelligibility prediction method for hearing loss that explicitly accounts for listener-specific degradations in frequency and temporal resolution. Our method tackles two significant issues: (1) simulating spectral and temporal deficits using auditory-inspired transformations, and (2) capturing their effects using STM analysis that reflects how these degradations alter perceptually relevant modulation patterns.

In contrast to conventional approaches based only on audiometric thresholds, our method captures how hearing loss distorts speech by analyzing STM representations. Using STM analysis and NCC-based similarity, the model learns how clean speech and SPIN differ in perceptually important modulation bands. Experiments on the Clarity Prediction Challenge corpus show that the proposed method provides more accurate predictions than the HASPI v2, particularly for the \textbf{Mild} group.

\label{sec:intro}

\clearpage
\bibliographystyle{IEEEtran}
\bibliography{refs25}

\begin{thebibliography}{10}
\providecommand{\url}[1]{#1}
\csname url@samestyle\endcsname
\providecommand{\newblock}{\relax}
\providecommand{\bibinfo}[2]{#2}
\providecommand{\BIBentrySTDinterwordspacing}{\spaceskip=0pt\relax}
\providecommand{\BIBentryALTinterwordstretchfactor}{4}
\providecommand{\BIBentryALTinterwordspacing}{\spaceskip=\fontdimen2\font plus
\BIBentryALTinterwordstretchfactor\fontdimen3\font minus \fontdimen4\font\relax}
\providecommand{\BIBforeignlanguage}[2]{{%
\expandafter\ifx\csname l@#1\endcsname\relax
\typeout{** WARNING: IEEEtran.bst: No hyphenation pattern has been}%
\typeout{** loaded for the language `#1'. Using the pattern for}%
\typeout{** the default language instead.}%
\else
\language=\csname l@#1\endcsname
\fi
#2}}
\providecommand{\BIBdecl}{\relax}
\BIBdecl

\bibitem{patrick2022hearing}
B.~J. Patrick, ``{Hearing Loss and Cognition: Using Behavioral Paradigms to Uncover the Learning Strategy Changes Following Noise-Induced Hearing Loss in Rodents},'' Master's thesis, The University of Western Ontario, 2022.

\bibitem{li2022prevalence}
W.~Li, Z.~Zhao, Z.~Lu, W.~Ruan, M.~Yang, and D.~Wang, ``{The Prevalence and Global Burden of Hearing Loss in 204 Countries and Territories, 1990--2019},'' \emph{Environ. Sci. Pollut. Res.}, pp. 1--8, 2022.

\bibitem{wilson2017global}
B.~S. Wilson, D.~L. Tucci, M.~H. Merson, and G.~M. O'Donoghue, ``{Global Hearing Health Care: New Findings and Perspectives},'' \emph{Lancet}, vol. 390, no. 10111, pp. 2503--2515, 2017.

\bibitem{miles2022measuring}
K.~Miles, T.~Beechey, V.~Best, and J.~B{\"u}chholz, ``{Measuring Speech Intelligibility and Hearing-Aid Benefit Using Everyday Conversational Sentences in Real-World Environments},'' \emph{Front. Neurosci.}, vol.~16, p. 789565, 2022.

\bibitem{ching1998speech}
T.~Y.~C. Ching, H.~Dillon, and D.~Byrne, ``{Speech Recognition of Hearing-Impaired Listeners: Predictions from Audibility and the Limited Role of High-Frequency Amplification},'' \emph{J. Acoust. Soc. Am.}, vol. 103, no.~2, pp. 1128--1140, 1998.

\bibitem{smoorenburg1992speech}
G.~F. Smoorenburg, ``{Speech Reception in Quiet and in Noisy Conditions by Individuals with Noise-Induced Hearing Loss in Relation to Their Tone Audiogram},'' \emph{J. Acoust. Soc. Am.}, vol.~91, no.~1, pp. 421--437, 1992.

\bibitem{pienkowski2017etiology}
M.~Pienkowski, ``{On the Etiology of Listening Difficulties in Noise Despite Clinically Normal Audiograms},'' \emph{Ear Hear.}, vol.~38, no.~2, pp. 135--148, 2017.

\bibitem{jin2010interrupted}
S.-H. Jin and P.~B. Nelson, ``{Interrupted Speech Perception: The Effects of Hearing Sensitivity and Frequency Resolution},'' \emph{J. Acoust. Soc. Am.}, vol. 128, no.~2, pp. 881--889, 2010.

\bibitem{porhun2021method}
M.~I. Porhun and M.~I. Vashkevich, ``{A Method for Simulation the Effect of the Reduced Frequency Resolution of the Ear in Patients with Sensorineural Hearing Loss},'' in \emph{Proc. Informatics}, vol.~18, no.~3, 2021, pp. 68--82.

\bibitem{haggard1985temporal}
M.~Haggard, ``{Temporal Patterning in Speech: The Implications of Temporal Resolution and Signal-Processing},'' in \emph{Proc. 11th Danavox Symp. on Hearing, Gamle Avern{\ae}s, Denmark}, 1985, pp. 215--237.

\bibitem{moore1992temporal}
B.~C.~J. Moore, M.~J. Shailer, and G.~P. Schooneveldt, ``{Temporal Modulation Transfer Functions for Band-Limited Noise in Subjects with Cochlear Hearing Loss},'' \emph{Br. J. Audiol.}, vol.~26, no.~4, pp. 229--237, 1992.

\bibitem{bacon1992modulation}
S.~P. Bacon and R.~M. Gleitman, ``{Modulation Detection in Subjects with Relatively Flat Hearing Losses},'' \emph{J. Speech Lang. Hear. Res.}, vol.~35, no.~3, pp. 642--653, 1992.

\bibitem{desloge2011temporal}
J.~G. Desloge, C.~M. Reed, L.~D. Braida, Z.~D. Perez, and L.~A. Delhorne, ``{Temporal Modulation Transfer Functions for Listeners with Real and Simulated Hearing Loss},'' \emph{J. Acoust. Soc. Am.}, vol. 129, no.~6, pp. 3884--3896, 2011.

\bibitem{chi1999spectro}
T.~Chi, Y.~Gao, M.~C. Guyton, P.~Ru, and S.~Shamma, ``{Spectro-Temporal Modulation Transfer Functions and Speech Intelligibility},'' \emph{J. Acoust. Soc. Am.}, vol. 106, no.~5, pp. 2719--2732, 1999.

\bibitem{edraki2020speech}
A.~Edraki, W.-Y. Chan, J.~Jensen, and D.~Fogerty, ``{Speech Intelligibility Prediction Using Spectro-Temporal Modulation Analysis},'' \emph{IEEE/ACM Trans. Audio Speech Lang. Process.}, vol.~29, pp. 210--225, 2020.

\bibitem{kates2021hearing}
J.~M. Kates and K.~H. Arehart, ``{The Hearing-Aid Speech Perception Index (HASPI) Version 2},'' \emph{Speech Commun.}, vol. 131, pp. 35--46, 2021.

\bibitem{kates2022overview}
------, ``{An Overview of the HASPI and HASQI Metrics for Predicting Speech Intelligibility and Speech Quality for Normal Hearing, Hearing Loss, and Hearing Aids},'' \emph{Hear. Res.}, vol. 426, p. 108608, 2022.

\bibitem{barker2024clarity}
J.~Barker, M.~A. Akeroyd, W.~Bailey, T.~J. Cox, J.~F. Culling, J.~Firth, S.~Graetzer, and G.~Naylor, ``{The 2nd Clarity Prediction Challenge: A Machine Learning Challenge for Hearing Aid Intelligibility Prediction},'' in \emph{Proc. ICASSP}, 2024, pp. 11\,551--11\,555.

\bibitem{nejime1997simulation}
Y.~Nejime and B.~C.~J. Moore, ``{Simulation of the Effect of Threshold Elevation and Loudness Recruitment Combined with Reduced Frequency Selectivity on the Intelligibility of Speech in Noise},'' \emph{J. Acoust. Soc. Am.}, vol. 102, no.~1, pp. 603--615, 1997.

\bibitem{moore1999loudness}
B.~C.~J. Moore, J.~I. Alcantara, M.~A. Stone, and B.~R. Glasberg, ``{Use of a Loudness Model for Hearing Aid Fitting: II. Hearing Aids with Multi-Channel Compression},'' \emph{Br. J. Audiol.}, vol.~33, no.~3, pp. 157--170, 1999.

\bibitem{moore1997model}
B.~C.~J. Moore and B.~R. Glasberg, ``{A Model of Loudness Perception Applied to Cochlear Hearing Loss},'' \emph{Auditory Neurosci.}, vol.~3, no.~3, pp. 289--311, 1997.

\bibitem{desloge2010speech}
J.~G. Desloge, C.~M. Reed, L.~D. Braida, Z.~D. Perez, and L.~A. Delhorne, ``{Speech Reception by Listeners with Real and Simulated Hearing Impairment: Effects of Continuous and Interrupted Noise},'' \emph{J. Acoust. Soc. Am.}, vol. 128, no.~1, pp. 342--359, 2010.

\bibitem{chi2005multiresolution}
T.~Chi, P.~Ru, and S.~A. Shamma, ``{Multiresolution Spectrotemporal Analysis of Complex Sounds},'' \emph{J. Acoust. Soc. Am.}, vol. 118, no.~2, pp. 887--906, 2005.

\bibitem{schadler2015separable}
M.~R. Sch{\"a}dler and B.~Kollmeier, ``{Separable Spectro-Temporal Gabor Filter Bank Features: Reducing the Complexity of Robust Features for Automatic Speech Recognition},'' \emph{J. Acoust. Soc. Am.}, vol. 137, no.~4, pp. 2047--2059, 2015.

\bibitem{ponsot2021mechanisms}
E.~Ponsot, L.~Varnet, N.~Wallaert, E.~Daoud, S.~A. Shamma, C.~Lorenzi, and P.~Neri, ``{Mechanisms of Spectrotemporal Modulation Detection for Normal- and Hearing-Impaired Listeners},'' \emph{Trends Hear.}, vol.~25, p. 2331216520978029, 2021.

\bibitem{bernstein2013spectrotemporal}
J.~G.~W. Bernstein, G.~Mehraei, S.~Shamma, F.~J. Gallun, S.~M. Theodoroff, and M.~R. Leek, ``{Spectrotemporal Modulation Sensitivity as a Predictor of Speech Intelligibility for Hearing-Impaired Listeners},'' \emph{J. Am. Acad. Audiol.}, vol.~24, no.~4, pp. 293--306, 2013.

\bibitem{edraki2021spectro}
A.~Edraki, W.-Y. Chan, J.~Jensen, and D.~Fogerty, ``{A Spectro-Temporal Glimpsing Index (STGI) for Speech Intelligibility Prediction},'' in \emph{Proc. Interspeech}, 2021, pp. 206--210.

\bibitem{dosovitskiy2020image}
A.~Dosovitskiy, L.~Beyer, A.~Kolesnikov, D.~Weissenborn, X.~Zhai, T.~Unterthiner, M.~Dehghani, M.~Minderer, G.~Heigold, S.~Gelly \emph{et~al.}, ``{An Image is Worth 16x16 Words: Transformers for Image Recognition at Scale},'' \emph{arXiv preprint arXiv:2010.11929}, 2020.

\bibitem{liu2025metacls}
X.~Liu, B.~Zhang, J.~Chen, B.~Shan, Z.~Han, and Y.~Ye, ``{MetaCLS: A Closer Look at [CLS] Token for Few-Shot Transformer},'' preprint, SSRN: 5201363, 2025. [Online]. Available: \url{https://ssrn.com/abstract=5201363}.

\bibitem{akiba2019optuna}
T.~Akiba, S.~Sano, T.~Yanase, T.~Ohta, and M.~Koyama, ``{Optuna: A Next-Generation Hyperparameter Optimization Framework},'' in \emph{Proc. ACM SIGKDD Int. Conf. Knowl. Discov. Data Min.}, 2019, pp. 2623--2631.

\end{thebibliography}







\end{document}